\documentclass[aps,twocolumn,prl,amssymb]{revtex4}
\usepackage{graphicx}
\usepackage{epsfig}
\begin{document}
\title{The Condition for Universality at Resonance and Direct Measurement of Pair Wavefunctions
Using rf Spectroscopy}
\author{Roberto B. Diener and Tin-Lun Ho}
\address{Department of Physics,  The Ohio State University,
Columbus, Ohio
43210}

\begin{abstract}
We show that when the Fermi energy of a Fermi gas is much smaller than the intrinsic energy width of a Feshbach resonance, the system behaves like a Fermi gas with a contact potential. This in turn implies universality at resonance, and large fermionic pairs in the strongly interacting regime. 
Recent experiments of JILA\cite{Jin} and MIT\cite{Ketterle} turn out to be deep inside the universal regime, which explains the perfect fit of these experiments by the BEC-BCS crossover theory with contact potential\cite{DienerHo}. We also show that rf spectrocopy can be used to map out the pair wavefunction directly. 
\end{abstract}

\maketitle

It is by now well known that a weakly interacting quantum gas can be made strongly interacting by tuning the system close to a scattering resonance,  where the s-wave scattering length becomes divergent. 
There are many ways to generate a scattering resonance.    The simplest one is to vary the pair potential on the microscopic scale.  This, however, is difficult to achieve. An alternate way, which leads to the recent explosion of activities, is to use Feshbach resonance. This resonance is achieved by Zeeman shifting the energy of a bound state in a closed channel to zero energy, thereby generating considerable resonance scattering for particles in the open channel.
In the last fourteen months, many remarkable properties of Fermi gases near a Feshbach resonance
were discovered --  universal interaction energy\cite{Duke,ENSenergy},  molecular condensates\cite{JinMBEC}, and the long sought condensation of fermion pairs. Evidence of such condensation was reported by C. Regal, M. Greiner, and D. S. Jin\cite{Jin} three months ago in $^{40}$K, and a month ago by Ketterle's group in $^{6}$Li\cite{Ketterle}. Evidence of superfluidity has also been observed recently by Grimm's~\cite{Grimm} and Thomas'~\cite{Thomas} groups.

As these phenomena emerge,  there are questions about whether they are general properties of scattering resonances, or specific properties of Feshbach resonances.  This question is sometimes phrased in a narrower context as whether single channel models and  two-channel (or resonance) models  have the same physics near resonance. The former  refers to models with two types of fermions interacting with a pair potential. The latter are those that describe open channel and closed channels physics explicitly.  Another important question is under what conditions are the properties of these models universal, i.e. independent of microscopic details. Moreover, if both types of models are universal,  do they belong to the same universality class?
These questions are not only of theoretical interest, but also of practical importance in interpreting 
current data and guiding future experiments. 

The purpose of this paper is to point out the condition for universal behavior (referred to as ``universality condition") for both single channel and two channel systems,  its implications on the ground states, and a simple method to measure the pair wave function directly. 
We shall show that {\bf (1)} If the Fermi energy is well within an intrinsic width of the resonance (defined later), then the interaction becomes $\delta$-function like, which guarantees universality at resonance.   {\bf (2)} There is only one universality class, given by the behavior of Fermi gases with contact interaction. 
{\bf (3)} Universality implies that the condensate must have large pairs (of the size of inter-particle spacing), where closed channel bound states play an insignificant role. 
 {\bf (4)} The  recent JILA and MIT experiments are deep in the universal regime.  This justifies the use of a single channel approach for  the JILA and MIT experiment\cite{DienerHo}, and explains the perfect agreement between the phase boundary in ref.\cite{Jin, Ketterle} and the $T_{c}$  predicted by the crossover theory with $\delta$-function potential\cite{DienerHo}.  {\bf (5)} Condensates with molecular rich (and hence small) pairs can be found in ``narrow" resonances.  
 {\bf (6)} RF spectroscopy can be used to map out the pair wavefunction across the resonance directly. 
 
The emergence of universality is most obvious for $\delta$-function potentials (in single channel models) with coupling constant $g=4\pi\hbar^2 a_{s}/M$, where $a_{s}$ is the s-wave scattering length and $M$ is the mass of the fermion.  At resonance, $a_{s}$ diverges. The only remaining length scale is then $n^{-1/3}$, where $n$ is the density. As a consequence, the thermodynamics of the system becomes universal\cite{unithermo}, i.e. independent of microscopic details.  It is clear that universal behaviors only  emerge when all other length scales become irrelevant.  However, all resonances  have an intrinsic length scale $r^{\ast}$, which is related to the width of the resonance in energy space.  
To be precise,  consider the scattering length near a Feshbach resonance, 
$a_{s} = a_{bg}( 1 - W/(B-B_{o}))$, where $a_{bg}$ is the background scattering length, $W$ is the width of the resonance, and $B_{o}$ is the location of the resonance.  Near resonance, it is sufficient to focus on the resonance term and write 
\begin{equation}
a_{s} = -\eta/\nu, \,\,\,\,\,\,\,\,\,  \nu = \mu_{co}(B-B_{o}), \,\,\,\,\,\,\, \eta = a_{bg}\mu_{co}W
\label{1} \end{equation}
where $\mu_{co}$ is the difference in magnetic moment between fermions in closed and open channel, and $\nu$ is the detuning energy.  
On can then define an intrinsic length scale $r^{\ast}$ or intrinsic width $(\Delta B)_{\rm in} $ in magnetic field as $\hbar^2/2Mr^{\ast} = \eta = r^{\ast}\mu_{co}(\Delta B)_{\rm in}$, or 
\begin{equation}
r^{\ast} = \frac{\hbar^2}{2M a_{bg} \mu_{co}W}, \,\,\,\,\,  (\Delta B)_{\rm in} = \frac{2M\mu_{co}(Wa_{bg})^2}{\hbar^2}. 
\label{rast} \end{equation}
In turn, one can write 
\begin{equation}
a_{s}/r^{\ast} =(\Delta B)_{\rm in}/(B_{o}-B)
\label{newa} \end{equation}
[For single channel systems where a bound state is generated when an energy parameters $u$ (such as the depth of the well) reaches a special value $u_{o}$, $a_{s}$ can still be written in the form eq.(\ref{1}) with $\nu=u_{o}-u$ and $\eta$ being some constant].  The presence of $r^{\ast}$ means the many body system has an additional dimensionless parameter $y=k_{F}r^{\ast}$, where
$k_{F}= (3\pi^2 n)^{1/3}$ is the Fermi wavevector.  We shall show that reduction to contact potential (hence universality at resonance) occurs when  
\begin{equation}
y \equiv k_{F}r^{\ast}<<1, \,\,\,\,\,   {\rm or} \,\,\,\,\,\, \mu_{co}(\Delta B)_{\rm in}/E_{F}  >>1. 
\label{unicon} \end{equation}

{\bf Other related works}:  Before proceeding, we would like to mention that eq.(\ref{unicon}) has also been recently identified by Eric Cornell as the condition for the equivalence between two channel and single channel models, (see later discussions)\cite{Cornell}.  While we agree with his conclusions, we point out eq.(\ref{unicon}) is in fact a (stronger) condition for universality, which implies channel equivalence.  
Such distinction is important because not all single channel models are universal, as we shall see later.  (The different potentials chosen in ref.\cite{DienerHo} happens to be all in the universal regime).  
In a recent prepint, G. Brunn\cite{Brunn} has also identified eq.(\ref{unicon}) as the condition for 
universal behavior when considering a normal state at $T=0$ without superfluid correlations.  The lack of superfluid correlation makes it difficult to connect to current experiments. 
During the writing of this paper, a preprint by De Palo et al.\cite{MariLiu} has appeared in which they come to similar conclusions for the single channel problem.  The present work differs from those mentioned above not only in perspective and approach, but also in the conceptual emphasis of underlying importance of universality,  which controls channel equivalence and the nature of the pairing state. In addition, we point out how rf spectroscopy can be used to probe the pair wavefunction in the universal regime. 

Before ending this section, we would like to paraphrase Cornell's arguments\cite{Cornell}, which we find illuminating. They should be appreciated together with the results of the explicit calculation presented later. Cornell notes that  {\bf (i)} For a two body system, when $|B-B_{o}|<< (\Delta B)_{\rm in}$,    
 or $a_{s}>>r^{\ast}$ in eq.(\ref{newa}),   the population of the closed channel is very small, and
the problem should therefore be single channel like~\cite{com1} . 
{\bf (ii)}   For a Fermi gas with density $n$, as $B$ approaches $B_{o}$,  two things can happen. Either one first enters the ``single channel region"  $|B-B_{o}|<< (\Delta B)$ (hence $a_{s}/r^{\ast}>>1$) and then the strongly interacting regime $n|a_{s}|^3 >>1$, or vice versa.   In the former case,  the system become single channel like before becoming strongly interacting.   {\bf (iii)} The condition for being single channel like can be obtained by replacing $\nu$ in eq.(\ref{newa}) by the Fermi energy $E_{F}$, which is the ``detuning" of the fermions at the Fermi surface. The ``single-channel" condition is then 
$\mu_{co}(\Delta B)_{\rm in}/E_{F}>>1$, 
which is eq.(\ref{unicon}).
We shall now prove the condition eq.(\ref{unicon}) using the two-channel model. 

{\bf Universality condition for two-channel models:}  We shall consider the resonance model\cite{two}
$H-\mu N$ $=$ $\sum_{\bf k, \sigma}( \epsilon_{\bf k} -\mu)a^{\dagger}_{{\bf k}, \sigma}a^{}_{{\bf k}, \sigma}$ $+ \sum_{\bf k} (\epsilon_{\bf k}/2 - 2\mu + \overline{\nu}) b^{\dagger}_{\bf k}b^{}_{\bf k} $
$ + \Omega^{-1/2} \alpha \sum_{\bf k, q} (b^{\dagger}_{\bf q} a^{}_{{\bf k} + {\bf q}/2, \uparrow} a^{}_{-{\bf k} + {\bf q}/2, \downarrow} + h.c.) $
where $a^{\dagger}_{{\bf k}\sigma}$ creates a fermion with momentum ${\bf k}$ and spin $\sigma$, 
($\sigma = \uparrow, \downarrow$);  $b^{\dagger}_{\bf k}$ creates a closed channel bound state with momentum ${\bf k}$, $\Omega$ is the volume of the system, and $\alpha$ is the (intensive) coupling between the closed channel bound state and the open channel scattering state. We shall ignore the background scattering length term since it is unessential for the resonance physics.  The quantity $\overline{\nu}$ is the ``bare" detuning which is related to the physical detuning by an infinite constant 
$\overline{\nu} = \nu + \frac{\alpha^2}{\Omega}\sum_{\bf k} \frac{1}{2\epsilon_{\bf k}}$ so as to cancel any unphysical ultra-violet divergences in the problem.
With this normalization, it is straightforward to solve for the two-body $T$-matrix and find the s-wave scattering length, which is $a_{s} = -\frac{M}{4\pi \hbar^2} \frac{\alpha^2}{\nu}$.

Next, we study the ground state properties of the resonance model using the BEC-BCS crossover theory\cite{single} where one assumes a condensation in the bound state $\langle b^{}_{\bf q=0}\rangle\equiv \Phi_{m} $  in the closed channel and a condensation of zero 
momentum pairs $\langle a_{{\bf k}, \uparrow}a_{-{\bf k}, \downarrow}\rangle$ in the open channel.
Standard approach then gives the mean field equation 
\begin{equation}
\frac{\nu-2\mu}{\alpha^2}  = \frac{1}{\Omega} \sum_{\bf k}\left( \frac{1}{2E_{\bf k}} - 
\frac{1}{2\epsilon_{\bf k}} \right) ,
\label{gap} \end{equation}
where $E_{\bf k} = \sqrt{ (\epsilon_{\bf k} -\mu)^2 + \Delta^2}$, and $\Delta$ is the gap parameter 
related to $\Phi_{m}$ as 
$\Delta = \alpha  \sqrt{n_{m}}$, $n_{m} = |\Phi_{m}|^2/\Omega$, 
where $n_{m}$ is the density of closed channel molecules.  The chemical potential is also constraint by the number equation
$n = 2n_{m} + \frac{1}{\Omega}\sum_{{\bf k},\sigma} \langle a^{\dagger}_{{\bf k},\sigma}a^{}_{{\bf k},\sigma}
\rangle$, or
\begin{equation}
n =\frac{2\Delta^2}{\alpha^2} + \frac{1}{\Omega}\sum_{\bf k} \left( 1 - \frac{ \epsilon_{\bf k} - \mu}
{E_{\bf k}} \right).
\label{number} \end{equation} 
Eq.(\ref{gap}) and (\ref{number}) together give $\mu$ and $\Delta$ (hence the 
molecular fraction $n_{m}/n$ in the closed channel) as a function of $n$ and the physical detuning $\nu$ for a given system $(\alpha)$. 

It is useful to write eq.(\ref{gap}) and (\ref{number}) in dimensionless form. Defining $q=k/k_{F}$,  $\tilde{\mu} = \mu/E_{F}$, $\tilde{\Delta}= \Delta/E_{F}$, where $E_F=\hbar^2k_{F}^2/2M$; and noting that $n =  k^{3}_{F}/3\pi^2$, it is easy to show that eq.(\ref{gap}) and (\ref{number}) can be written in dimensionless form  $y (\tilde{\nu}-2\tilde{\mu}) = F_{1}(\tilde{\mu}, \tilde{\nu})$, 
$(4/3\pi^2) = y\tilde{\Delta}^2 +  F_{2}(\tilde{\mu}, \tilde{\Delta})$, where $F_{1}$ and $F_{2}$ are functions of  $\tilde{\mu}$ and $\tilde{\Delta}$\cite{F1F2}.  Microscopic details are characterized by $y$ only.

\begin{figure}[t]
\epsfxsize=2.5in
\epsffile{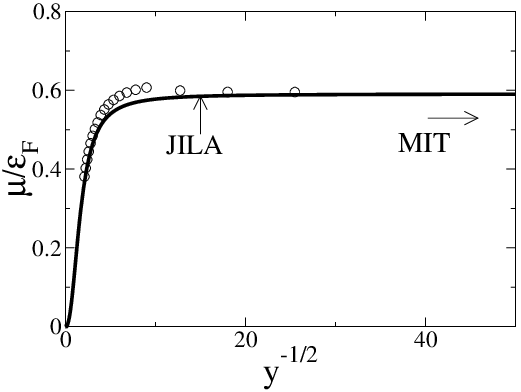}
\caption{Chemical potential on resonance $(\nu=0)$ as a function of $y^{-1/2}$, $y=k_{F}r^{\ast}$. The solid line and circles are results of two-channel and single channel calculations respectively. Both models exhibit universality (independent of microscopic detail $y$) for $y^{-1/2}>10$. 
The corresponding values in the JILA\cite{Jin} and MIT\cite{Ketterle} experiments are $y^{-1/2}=15$ and $80$, which are inside and deep inside the universal regime.}
\end{figure}

The emergence of universality is best illustrated in figure 1 where we have plotted $\tilde{\mu}$ at resonance $\tilde{\nu}=0$ as a function of $1/\sqrt{y}$.  (There is no particular significant in this choice
of plotting variable except it gives a clear display of approach to universality). 
We see that as $y^{-1/2}$ exceeds beyond $10$, $\tilde{\mu}$ becomes independent of $y$ (hence microscopic details). Moreover, $\tilde{\mu}$ saturates at 0.59, the same value of the single channel $\delta$-function potential.    At the same time, the fraction of close channel molecules $n_{m}/n= |\Phi_{m}|^2/N$,  ($n_{m} = \Delta^2/ \alpha^2$) drops rapidly as $y^{-1/2}$ increases, as shown in figure 2.  {\em This shows that the molecular component is insignificant in the universal regime. The size of the fermion pair must therefore be of order $n^{-1/3}$.} 
The behavior of $\tilde{\mu}$ across resonance (i.e. plotted for different $1/(k_{F} a_{s})$) for different $\alpha$ is shown in figure 3.  For $y^{-1/2}>10$, all $\tilde{\mu}$ coincide, and is precisely that given by  single channel systems with $\delta$-function potential, demonstrating the universality of the contact potential in the wide resonance regime.  

The reduction to single channel $\delta$-function can in fact be proven simply as follows.  By noting that the term $\nu/\alpha^2$ in eq.(\ref{gap}) is simply $-1/g$, where 
$g= 4\pi\hbar^2 a_{s}/M$, eqs.(\ref{gap}) and (\ref{number})  reduce to 
\begin{equation}
-\frac{1}{g} = \frac{1}{\Omega} \sum_{\bf k}\left( \frac{1}{2E_{\bf k}} - 
\frac{1}{2\epsilon_{\bf k}} \right) ,
\label{scgap} \end{equation}
\begin{equation}
n = \frac{1}{\Omega}\sum_{\bf k} \left( 1 - \frac{ \epsilon_{\bf k} - \mu}
{E_{\bf k}} \right).
\label{scnumber} \end{equation} 
in the limit of large $\alpha^2$ since $\mu/\alpha^2$ and $\Delta^2/\alpha^2$ can be dropped from eqs.(\ref{gap}) and (\ref{number}) respectively.  Eq.(\ref{scgap}) and (\ref{scnumber}) are precisely 
the gap equation and number equation for the single channel systems with contact potential.  

\begin{figure}
\epsfxsize=2.5in
\epsffile{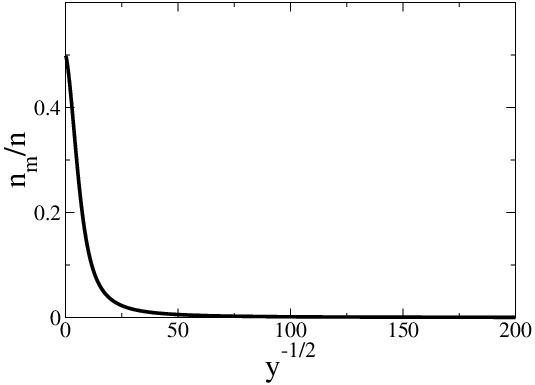}
\caption{The fraction of closed channel molecules on resonance $\nu=0$ as a function of 
$y^{-1/2}$, $y=k_{F}r^{\ast}$. One sees that the fraction of closed channel molecules is insignificant in the universal regime.}
\end{figure}

{\bf Single channel models:} It is important to note that {\em not all single channel models are universal.}
This is also pointed out in ref.~\cite{MariLiu}. To show that the condition 
for universality is also given by eq.(\ref{unicon}), we consider 
a square well potential with a barrier: $V(r) = -u_{o}<0$, $r<r_{o}$; $V(r)= u_{1}>0$ for $r_{o}>r>r_{1}$, and $V(r)=0$ for $r>r_{1}$.  As the parameters 
 $u_{o}$ and $u_{1}$ are varied, different values of $y= k_{F}r^{\ast}$ are generated.  In figure 1, we have also displayed (with circles) the chemical potential along a path in parameter space which gives a monotonically increasing $y^{-1}$\cite{path}.  (There are many such paths). One sees from fig.1 that the chemical potential again reaches a constant when $y^{-1/2}  {\gtrsim} 10$, which is a sign of universality. 
 
{\bf Relation to current experiments:}
In the JILA experiment\cite{Jin},  one has $n=10^{13}{\rm cm}^{-3}$,  $W=8$ G, 
$a_{bg}=170a_{B}$, $\mu_{co}\sim 2\mu_{B}$, where $a_{B}$ and $\mu_{B}$ are the Bohr radius and Bohr magneton.  Since $\eta = \mu_{co}W a_{bg}$, we find from eq.(\ref{rast}) that 
$y^{-1/2}= 15$, (or $y=k_{F}r^{\ast}=0.004$) which is inside the universal region as shown in figure 1.  
In the MIT experiment\cite{Ketterle}, one has $k_{F}^{-1}=2000a_{B}$, $W=180$ G, $a_{bg}= -2000a_{B}$, $\mu_{co}\sim 2\mu_{B}$, which implies $y^{-1/2}\sim 80$, (or $y= 1.5 \times 10^{-4}$), which is deep inside the universal single channel regime.
This also explains the surprisingly good agreement between the prediction of crossover theory using a $\delta$-function potential with the experiments in  ref.\cite{Jin} and \cite{Ketterle} as pointed out in ref.\cite{DienerHo}. 

\begin{figure}[b]
\epsfxsize=2.5in
\epsffile{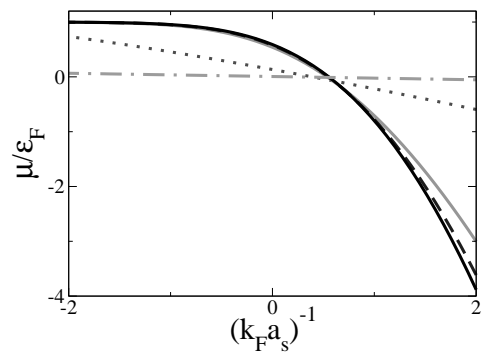}
\caption{The behavior of chemical potential across resonance for different values of $y^{-1/2}$: 
0.25 (dot-dashed), 1 (dotted), 5 (solid grey), 10 (dashed), 25 (solid black).
 For $y>10$, all curves coincide with that of single channel contact potential.}
\end{figure}
\begin{figure}[t]
\epsfxsize=2.5in
\epsffile{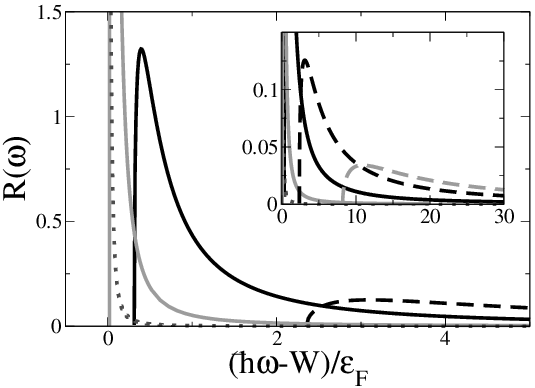}
\caption{Transition rate in rf spectroscopy experiments (in arbitrary units) as a function of frequency.  The values of $(k_{F}a_{s})^{-1}$ for different curves are: 
2 (dashed grey), 1 (dashed black), 0 (solid black), -1 (solid grey), -2 (dotted).}
\end{figure}
{\bf Direct Measurement of Pair Wavefunction:} We conclude by pointing out that rf spectroscopy can be used to determine the density profile of the pair wavefunction directly. It will therefore be a useful tool to explore the properties in the universal regime\cite{other}.  
Consider exciting a fermion $a_{\uparrow}$ to a different atomic state $c$. For example,  $a_{\uparrow}$ and $c$ can be the hyperfine states $F_{z}=-5/2$ and $-7/2$ respectively. 
Let $W$ be the energy difference between $a_{\uparrow}$ and $c$.  
The inclusion of the $c$ particle (of the same mass) and the rf field will introduce an additional term
$\sum_{\bf k}\epsilon_{k} c^{\dagger}_{\bf k} c^{}_{\bf k} $ to the unperturbed Hamiltonian and a time dependent perturbation $V=  \lambda \sum_{\bf k}
(c^{\dagger}_{\bf k}a^{}_{{\bf k}, \uparrow} e^{-i\omega t} + h.c.)$. 
The rate of transition from ground state to the $c$ state is given by the Fermi Golden Rule, ${\cal R}  = \frac{2\pi}{\hbar} \sum_{f} \left| \langle f | \lambda c^{\dagger}_{\bf k} a^{}_{{\bf k}, \uparrow}|G\rangle\right|^2 \delta( {\cal E}_{f} - {\cal E}_{i} -\hbar \omega)$, where $|G\rangle$ and $|f\rangle$ are the 
ground state and the excited states of the system, 
with energies ${\cal E}_{i}$ and ${\cal E}_{f}$ respectively.   Considering an initial fermion ground state with $2N$ particles with energy $E^{G}_{2N}$, we have ${\cal E}_{i}=E^{G}_{2N}$. 
Since the excited state consists of a Bogoliubov particle with energy $E_{\bf k}=\sqrt{(\epsilon_{\bf k}-\mu)^2 + \Delta_{\bf k}^2}$ on top of a ground state with energy $E^{G}_{2N-1}$ and a  particle $c^{\dagger}_{\bf k}$ with energy $\epsilon_{\bf k}+ W$, we have 
${\cal R} (\omega) = \frac{2\pi \lambda^2}{\hbar} \frac{1}{\Omega}\sum_{\bf k} v_{\bf k}^2 \delta(E_{\bf k} + W + \epsilon_{\bf k}- \mu - \hbar \omega)$, or 
\begin{equation}
{\cal R} (\omega) = \frac{2\pi \lambda^2}{\hbar} \frac{1}{\Omega} D(\epsilon^{\ast})\left|  \frac{ v^{2}_{\bf k}}
{1 + \partial E_{\bf k}/\partial \epsilon_{\bf k}}\right|_{\epsilon^{\ast}}
\label{RRR} \end{equation}
where $v_{\bf k} = \sqrt{[1 -(\epsilon_{\bf k} - \mu)/E_{\bf k})]/2}$ is  the coherence factor, 
$D(\epsilon)= (2\pi)^{-3} d^{3}k/{\rm d}\epsilon_{\bf k}$ is the density of state,  
and $\epsilon^{\ast}$ is the solution of the equation 
$\hbar \omega - W =  \sqrt{ (\epsilon^{\ast} - \mu)^2 + \Delta^2}  + \epsilon^{\ast}-\mu$. 
For $\delta$-function potentials, $\Delta_{\bf k}$ is a constant, we then have $1 + \partial E_{\bf k}/\partial \epsilon_{\bf k}= 1- v^{2}_{\bf k} \equiv  u_{\bf k}^2$.  Eq.(\ref{RRR}) then reduces to 
 \begin{equation}
{\cal R}(\omega) = \frac{2\pi \lambda^2}{\hbar} D(\epsilon^{\ast})\left|  v_{\bf k}/u_{\bf k} \right|^{2}_{\epsilon^{\ast}}. 
\label{finalR} \end{equation}
Since $v_{\bf k}/u_{\bf k}$ is the Fourier transform of the pair wavefunction apart from a normalization constant\cite{single}, and since $D(\epsilon) \propto \sqrt{\epsilon}$, one can therefore extract directly from the 
signal of rf spectroscopy (eq.(\ref{finalR})) the coefficient $v_{\bf k}/u_{\bf k}$ and reconstruct the real space wavefunction.  The behavior of eq.(\ref{finalR}) for fermions interacting with a $\delta$-function potential in BCS, resonance, and BEC regimes are shown in figure 4 for the case of $W>0$. The case of $W<0$ can be obtained by flipping about the vertical axis, i.e. changing $\hbar\omega - W$ to $W-\hbar \omega$. Note that the shape of the curve in the molecular regime is consistent with the shape observed
in ref.\cite{rfJILA} (which has $W<0$). 

We thank Hans Peter Buchler for a discussions on rf spectroscopy. 
This work is supported by NASA GRANT-NAG8-1765  and NSF Grant DMR-0109255.


\begin{thebibliography}{99}
\bibitem{com1} The situation is of course less clear 
for a many-body system. Nevertheless, one can proceed to see where it leads.
\bibitem{path} The first 16 points (for smaller $y^{-1/2}$) correspond to a square well with $k_{F}r_0 = 0.05$ and a bump with $r_1/r_0 = 1.2$ and decreasing height $u_1 = 15 u_{0}^{\rm min} ...  0$.  Here $u_0^{\rm min} = \frac{\hbar^2\pi^2}{4mr_0^2}$ is the depth of a well of radius $r_0$ for which a bound state appears.  In this calculation, the depth $u_0$ is adjusted to be in resonance.
The last 3 points correspond to a square well (without a bump) with a radius such that $k_{F}r_0 = 0.05/2^n$, with $n=1,2,3$.
\end{thebibliography}
\end{document}